\documentclass[aps,pre,twocolumn,amsfonts,amssymb,floatfix,nofootinbib,notitlepage]{revtex4-1}
\usepackage{amssymb}
\usepackage{amsmath}
\usepackage{graphicx,color}
\usepackage{bbm}
\usepackage{multirow}
\usepackage[colorlinks,linkcolor=blue,urlcolor=blue]{hyperref}

\allowdisplaybreaks

\newcommand{\beq}{\begin{equation}}
\newcommand{\eeq}{\end{equation}}
\newcommand{\beqnn}{\begin{equation*}}
\newcommand{\eeqnn}{\end{equation*}}
\newcommand{\bea}{\begin{eqnarray}}
\newcommand{\eea}{\end{eqnarray}}
\newcommand{\beann}{\begin{eqnarray*}}
\newcommand{\eeann}{\end{eqnarray*}}

\begin{document}

\title{On the protocol dependence of plasticity in ultra-stable amorphous solids}

\author{Edan Lerner$^1$, Itamar Procaccia$^2$, Corrado Rainone$^2$ and Murari Singh$^2$}
\altaffiliation{Laboratoire Charles Coulomb (L2C), University of Montpellier, CNRS, Montpellier, France.}
\affiliation{$^1$Institute for Theoretical Physics, University of Amsterdam, Science Park 904, 1098 XH Amsterdam, The Netherlands.\\ $^2$ Department of Chemical Physics, the Weizmann Institute of Science, Rehovot 76100, Israel.}

\begin{abstract}
While perfect crystals may exhibit a purely elastic response to shear all the way to yielding, the response of amorphous solids is punctuated by plastic events. The prevalence of this plasticity depends on the number of particles $N$ of the system, with the average strain interval before the first plastic event, $\overline{\Delta\gamma}$, scaling like $N^\alpha$ with $\alpha$ negative: larger samples are more susceptible to plasticity due to more numerous disorder-induced soft spots. In this paper we examine this scaling relation in ultra-stable glasses prepared with the Swap Monte Carlo algorithm, with regard to the possibility of protocol-dependent scaling exponent, which would also imply a protocol dependence in the distribution of local yield stresses in the glass. We show that, while a superficial analysis seems to corroborate this hypothesis, this is only a pre-asymptotic effect and in fact our data can be well explained by a simple model wherein such protocol dependence is absent.
\end{abstract}

\maketitle

\section{Introduction}
Besides their obvious difference in structure, amorphous solids and crystals present qualitative differences in their response to quasi-static mechanical loading. To fix ideas, let us focus on athermal solids, and consider a simple shear protocol wherein only the component $\gamma_{xy}$ of the strain tensor is non-zero. A perfect crystal will generally show a linear response~\cite{AshcroftMermin}, with a shear stress $\sigma$ proportional to the strain $\gamma$, all the way until the yield strain $\gamma_Y$ of the solid whereupon the crystal will fail and plastic flow will be initiated.
Not so for amorphous solids, such as glasses (colloidal, metallic, molecular), foams, and pastes. In these materials even the linear-response regime is substantially plastic~\cite{11HKLP}: the response is punctuated by sharp stress drops, giving the stress-strain curve a serrated appearance~\cite{04VBB,04ML,05DA,06TLB,06ML,09LP,11RTV}.
The density of this serration is determined by the size of the system which is parameterized by the number of particles $N$. Specifically, denote as $\overline{\Delta\gamma}$  the average strain interval after which a virgin material undergoes its first plastic event. Here the overline denotes an average over different virgin realizations. This quantity is found to scale like  $N^{\alpha}$, with $\alpha \simeq -0.6$~\cite{KLP10b}; this is intuitively related to the greater ease, in larger systems, of finding soft spots, thereby making large samples show plasticity at smaller strains; in the thermodynamic limit, plastic events occur with arbitrarily small strains \cite{11HKLP}. This  property of amorphous and disordered systems is presently referred to as \emph{marginal stability}~\cite{15MW}. If one keeps straining the material up to its yielding point $\gamma_Y$, the stress is seen to reach  a plastic regime where on the average the stress remains constant due to a balance between elastic increases and  plastic drops~\cite{09LP}.
These observations are general to a variety of model systems.

An issue of concern in the context of numerical simulations of this phenomenology had been that
molecular dynamics is limited in providing deeply supercooled equilibrated liquids. Thus typically amorphous solids at very low temperature could only be prepared by quenching from melts equilibrated at temperatures that are just inside the supercooled branch.
This state of affairs has changed in the last years, with a new flourish of efforts aimed at designing enhanced protocols for glass preparation, such as vapor deposition~\cite{Sw07,LEDP13} and Swap Monte Carlo (SMC)~\cite{GP01,06FLV,GKPP15,BCCNOY17,NBC17}. In particular in ref.~\cite{OBBRT18} the authors study the stress response of ultra-stable glasses prepared from melts equilibrated at exceptionally low temperatures with SMC, which greatly expands on the picture reported above. In particular, ultra-stable glasses seem to show a crystal-like response to loading, with a pre-yield phase almost completely devoid of plasticity. Concerning mechanical yield, the typical stress peak turns into a macroscopic stress drop, associated with the nucleation of a shear band and a sudden release in stress, enabling the authors to interpret their results in terms of a supercooling-induced transition from ductile to brittle yield~\cite{OBBRT18}, which they also back up with a description in terms of a mean-field model of plasticity. An equivalent point of view has been advanced in refs.~\cite{PGW18,WPGLW18}, suggesting a dramatic change in the nature of both plasticity and yield in well-annealed and ultra-stable amorphous materials.

The two main signatures of this putative transition from ductile to brittle behavior (a macroscopic stress jump and a depletion of plastic events in the pre-yield phase) can both be understood in terms of the local mechanical properties. These are modelled by partitioning a system into cells and asking what is the minimal strain for which a particular cell exhibits a plastic event triggering
 an instability in the material. Denoting this local threshold as $x$ one is interested in the distribution of thresholds $P(x)$, whose tail as $x\to 0$ is assumed to be characterized by a critical exponent $\theta$, $P(x) \simeq x^\theta$ when $x\to0$~\cite{15MW,LW16} (known as a pseudogap). In particular, it is demonstrated in~\cite{OBBRT18} that a more peaked $P(x)$, used to represent a better annealed glass within a simple elasto-plastic model, does reproduce the larger stress peak. Besides, the strain interval exponent $\alpha$ can be related to the pseudogap exponent $\theta$ via extreme value statistical arguments~\cite{KLP10b}.

Given the discussion above, it is therefore pertinent to investigate the possibility that the exponent $\theta$ depend on the preparation protocol, as indeed argued, and supported by some numerics, in refs.~\cite{OBBRT18,PGW18}. In this paper, we perform numerical simulations of SMC glasses and extract the scaling exponent $\alpha$ in a wide range of preparation temperatures, as a proxy for $\theta$. We will show that the measurement, while seemingly indicating a protocol dependent $\alpha$ for small (but still fairly large in terms of computational demands) system sizes, is actually affected by strong finite size effects, which can be interpreted in terms of a simple model for $P(x)$ characterized by a universal, i.e. protocol-independent, pseudogap exponent. These finite size effects can prevent the statistical arguments of refs.~\cite{KLP10b,LLRW14,LGRW15} from holding unless the system size is sufficiently large, the sizes required becoming larger and larger as the preparation temperature is reduced. Our conclusion is that the hypothesis of a protocol dependent pseudogap exponent cannot be supported by our data or those reported in refs.~\cite{OBBRT18,PGW18}.

\section{Materials and methods}
 We simulate in two dimensions $N$ point particles having equal mass $m=1.0$ with polydisperse sizes drawn from a probability distribution $P(\sigma)\sim \frac{1}{\sigma^3}$, in a range between small ($s$) and large ($l$) particles such that $\sigma_s/\sigma_l = 0.45$ and $\overline{\sigma} = 1.0$. Particles interact via a purely repulsive soft inverse power-law (IPL) potential given by the expression
  \begin{equation}
U(r_{ij})= \epsilon\left(\frac{\sigma_{ij}}{r_{ij}}\right)^{12} + C_0  + C_2 \left(\frac{r_{ij}}{\sigma_{ij}}\right)^2 +  C_4 \left(\frac{r_{ij}}{\sigma_{ij}}\right)^4,
  \end{equation}
  where the cross diameter $\sigma_{ij}$ follows the non-additive rule $\sigma_{ij}=\frac{\sigma_i+\sigma_j}{2}(1-0.2|\sigma_i-\sigma_j|)$
  to enhance glass-forming ability~\cite{NBC17,BCCNOY17}. The parameters $C_0, C_2$ and $C_4$ are chosen such that the potential and its first and second derivatives all vanish at a cutoff $R_{cut}=1.25\sigma_{ij}$. The potential is therefore the same (but in 2$d$) as the one used in ref.~\cite{OBBRT18}. To aid in the interpretation of the results, we also estimated its Mode Coupling Transition (MCT) temperature using the procedure employed in ref.~\cite{BCNO16}, producing an estimate $T_{MCT} = 0.274\pm 0.001$ (see appendix~\ref{app:MCT}).
  The reduced units for mass, length, energy and time have been taken as
  $m$, $\overline{\sigma}$, $\epsilon$ and $\overline{\sigma}\sqrt{m/\epsilon}$ respectively.

To prepare glasses, we employ a simple protocol wherein a high-temperature melt is equilibrated at a fixed temperature $T_g$, and then quenched out of equilibrium to form a glass. Within such a setting, the protocol dependence can be simply ``parametrized'', following~\cite{RUYZ15,Rainone17}, in terms of $T_g$, i.e. the last temperature whereupon the system was at equilibrium. This temperature was first introduced by Tool in ref.~\cite{Tool46} and therein denoted as $T_f$ (i.e. the fictive temperature); here we denote it as $T_g$ following~\cite{RUYZ15,Rainone17}.
We therefore start with a random configuration generated at density $\rho=1.1$ and then we equilibrate it at $T=T_g$ using the Swap Monte Carlo~\cite{GP01,GKPP15,BCCNOY17,NBC17} algorithm. This algorithm consists of standard Monte Carlo moves augmented by swap moves of particles of different sizes, leading to a dramatic speed-up of relaxation with respect to standard Molecular Dynamics (MD)~\cite{GP01}. After equilibrating at given $T_g$, we then employ standard MD to cool down the system with a rate of $\dot{T}=1\times10^{-5}$ down to $T=10^{-6}$, and then we finally quench it instantly to the nearest inherent minimum at $T=0$ using conjugate gradient minimization. We repeat the above procedure, e.g. 1000 times, to produce an ensemble of athermal amorphous solids. These solids are then strained with a standard AQS protocol~\cite{06ML}. We explore the temperature range from $T_g=0.5$ to $T_g=0.08$ to equilibrate the system using SMC, and we study system sizes of $N=1000, 2000, 4000, 10000$ and $N=40000$ particles.

\section{Preliminary results: system-size scaling of plasticity}
\begin{figure}[tb!]
 \begin{center}
  \includegraphics[width=0.45\textwidth]{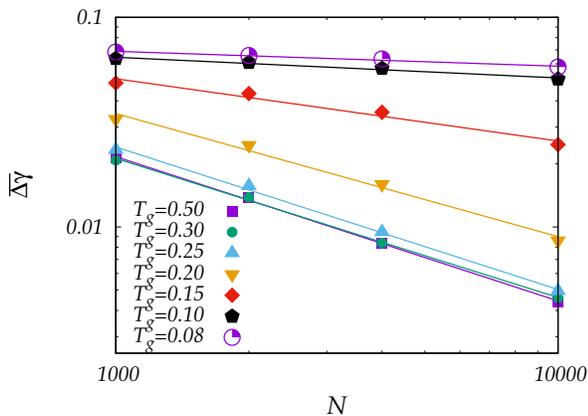}
  \caption{System size dependence of the average $\overline{\Delta\gamma}$ interval before the first plastic event occurs. For each $T_g$ The expected scaling $N^{\alpha}$ is found, but the $\alpha$ exponent is now dependent on the preparation temperature $T_g$.\label{fig:deltagvsNpre}}
 \end{center}
\end{figure}

As these athermal configurations (or equivalently, amorphous minima) are strained, then at a certain value $\Delta\gamma$ of the accumulated strain the first plastic event will occur. We measure the value of $\Delta\gamma$ for each configuration and then compute its average $\overline{\Delta\gamma}$ over all available minima, for a fixed system size $N$. We then repeat the process for all sizes from $N=1000$ and $N=10000$, in order to obtain the system size dependence of $\Delta\gamma$ for each given preparation temperature $T_g$.
The results are presented in Fig.~\ref{fig:deltagvsNpre}. For each $T_g$, $\overline{\Delta\gamma}$ exhibits power-law scaling, as expected and previously reported~\cite{KLP10b}. And thanks to the freedom of choice of $T_g$ conferred to our numerics by the SMC algorithm, we are able to reveal that the scaling exponent $\alpha$ is also, apparently, dependent on $T_g$.

The degree of plasticity of a metastable glass therefore appears strongly protocol-dependent: glasses with a lower $T_g$, corresponding therefore to a longer preparation protocol, are found to be a lot less plastic than glasses prepared at higher temperatures, in agreement with the results of refs.~\cite{OBBRT18,WPGLW18}. In the following we are going to show that this is not actually the case and that the effect is actually due to strong finite-size effects, which can however be hard to detect.

\section{Extreme value theory}
We start from the argument of refs.~\cite{KLP10b,LLRW14,LGRW15} linking the $\alpha$ and $\theta$ exponents. If one assumes, reasonably, that the first plastic event is originated in the softest region of the material, then one must study the statistics of the minimal number extracted from the distribution $P(x)$. Given that $P(x)$ has a pseudo-gapped form and a compact support (because of mechanical stability, $x$ cannot be negative) it then follows from extreme value theory~\cite{Gumbel2012} the following relation between the two exponents
\beq
\alpha = -\frac{1}{1+\theta}
\label{eq:alphavtheta}
\eeq
and that the pdf of the scaling variable $\Delta\gamma N^{-\alpha}$ must abide by a Weibull~\cite{Weibull1939} law; the same reasoning is followed in ref.~\cite{WPGLW18}. In ref.~\cite{OBBRT18}, the $\theta$ exponent is instead linked (and measured through) the system size scaling of the average avalanche size, but also in that case the argument makes use of extreme value statistics to estimate the density of avalanches~\cite{LGRW15}.
Eq.~\eqref{eq:alphavtheta}, taken together with the results of Fig.~\ref{fig:deltagvsNpre}, would imply a protocol dependence of the exponent $\theta$, with it increasing when $T_g$ is lowered, therefore leading to a depletion of soft excitations in well-annealed glasses.

\begin{figure}[tb!]
 \begin{center}
  \includegraphics[width=0.45\textwidth]{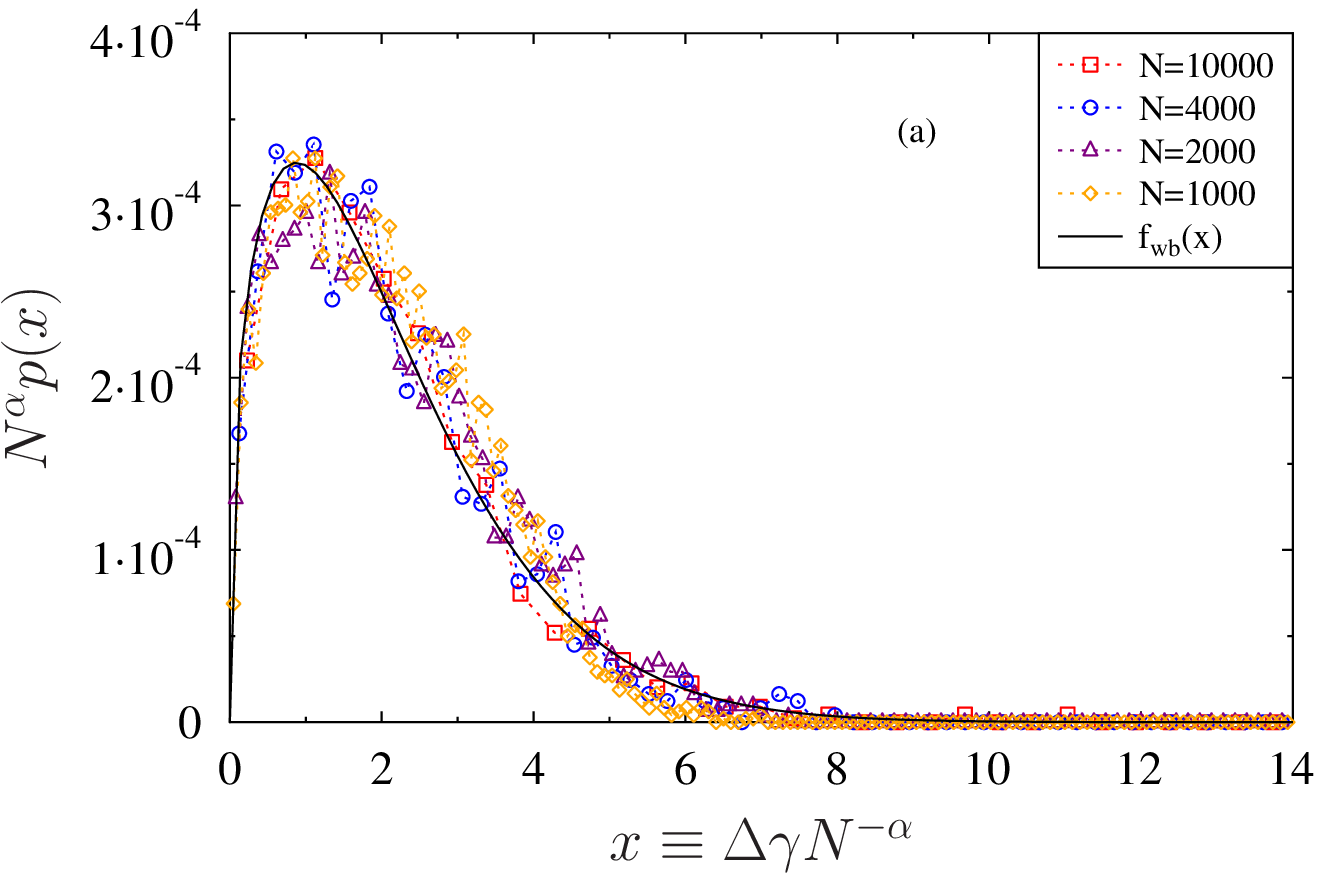}
  \includegraphics[width=0.45\textwidth]{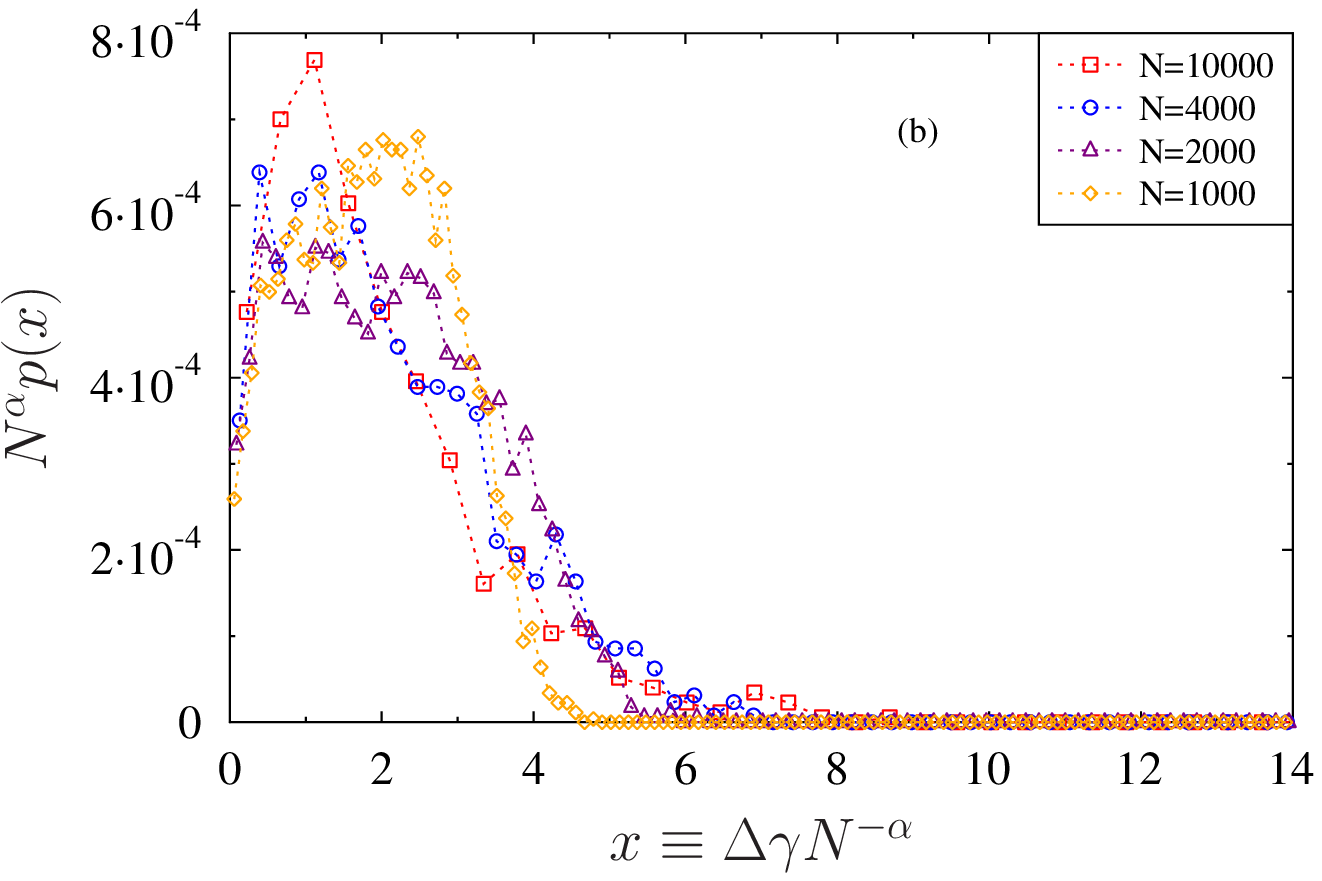}
  \includegraphics[width=0.45\textwidth]{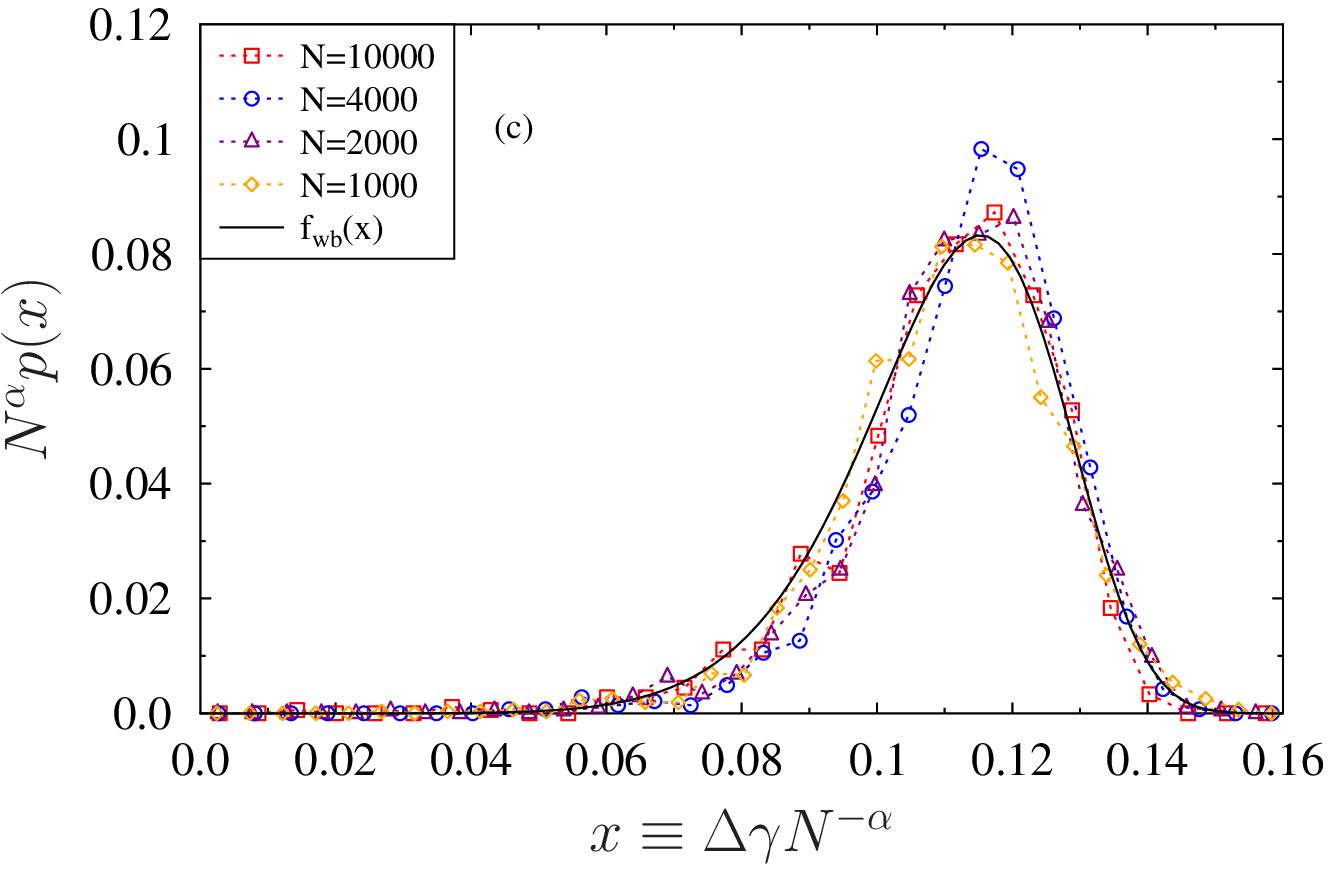}
  \caption{The rescaled pdf of $\Delta\gamma$ for $T_g=0.3$ (a), $T_g=0.2$ (b) and $T_g=0.08$ (c). Data collapse takes place only at "high" (i.e. larger than $T_{MCT}$) and "low" (i.e. deeply supercooled) temperatures, but not in the intermediate regime. \label{fig:Weibulls}}
 \end{center}
\end{figure}

This is however not the case. A way to reveal this is to look at not just at the average of $\Delta\gamma$, but at its full pdf. If the argument reported above holds, one should be able to collapse the data for all system sizes (at a single preparation temperature) on a single master curve of the Weibull form. In Fig.~\ref{fig:Weibulls} we report the pdfs of $\Delta\gamma$ for $T_g=0.3$ (i.e. slightly above $T_{MCT}$), $T_g =0.2$, and $0.08$ (i.e. in the deeply supercooled regime).
 Remarkably, data collapse is satisfactory, and the Weibull form is able to fit the data, only for $T_g=0.3$ and $T_g=0.08$, but not at the intermediate temperature $T_g=0.2$. This requires an
 explanation which is provided next.

\section{A toy model}
We resolve this quandary with a simple toy model. We obviously assume $P(x)$ to have a pseudo-gapped form, with $P(x) = c(T_g)x^\theta$ for $x\to 0$ (see appendix~\ref{app:toymodel} for more details). However, we assume the critical exponent $\theta$ to be universal, with all protocol dependence limited to the prefactor $c(T_g)$, which we assume to decrease as $T_g$ is lowered. These assumptions are reasonable in light of some measurements of the $P(x)$ in model glasses~\cite{BLHGFVP18}, and measurements of the Density of States of plastic, quasi-localized harmonic modes (on these modes, see for example~\cite{LDB16,KBL18} and references therein) of SMC glasses, wherein only the DOS' prefactor is found to depend on the degree of annealing~\cite{18LB,WNGBSF18}. We choose $\theta=2/3$ which implies $\alpha=-3/5=0.6$, but this choice is arbitrary and not relevant to the discussion that follows.

\begin{figure}
 \begin{center}
  \includegraphics[width=0.45\textwidth]{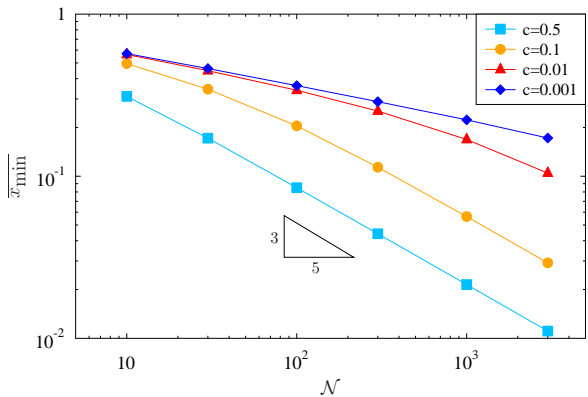}
  \caption{As $c$ decreases (and therefore the glass is better annealed), the asymptotic scaling $\overline{x_{\textrm min}} \propto \mathcal{N}^{-3/5}$ is pushed to larger and larger system sizes, producing an apparent change in the scaling exponent $\alpha$, which is however only a finite-size effect.\label{fig:x_min}}
 \end{center}
\end{figure}

This model $P(x)$ falls again under the hypotheses of extreme value statistics theorems~\cite{Gumbel2012} which prescribe a Weibull form for the pdf of the minimal $x$ taken from an asymptotically large collection of numbers drawn from $P(x)$, and a scaling law $\overline{x_{\textrm min}} \propto \mathcal{N}^{-3/5}$ of its average with the collection's size $\mathcal{N}$, independently of the value of $c(T_g)$ and therefore of the protocol.
This is not however the case in the pre-asymptotic regime, as we show in Fig.~\ref{fig:x_min}.
When $c$ is reasonably large (therefore corresponding to high $T_g$ and a poorly annealed glass), one is immediately able to see the proper asymptotic scaling, even at the smallest sample sizes we simulate. However, when $c$ is decreased, the asymptotic regime is pushed to larger sizes, and eventually ends up outside the observation window. This finite size effect will lead to an apparent change in the scaling exponent despite $\theta$ here being protocol-independent by construction, unless one takes care to simulate larger and larger systems as the preparation temperature is decreased.

We remark that unless this requirement is met, the effect can be hard to detect. With a constant (i.e. $T_g$-independent) range of sizes, such as the one in Fig.~\ref{fig:deltagvsNpre}, it is easy to be deceived into inferring that the scaling exponent is indeed changing: at high temperature the right asymptotic scaling is found, whilst at very low temperature the asymptotic regime is so far away that a power-law can anyway be reliably fitted to the curves. Only for intermediate temperatures one will be able to observe deviations from power-law scaling due to the crossover between pre-asymptotic and asymptotic regime taking place within the observation window, producing a breakdown of data collapse such as the one we report in Fig.~\ref{fig:Weibulls}, panel $(b)$; the effect can however still not be immediately apparent by simple inspection of the scaling plots of the averages.

\begin{figure}[tb!]
 \begin{center}
  \includegraphics[width=0.45\textwidth]{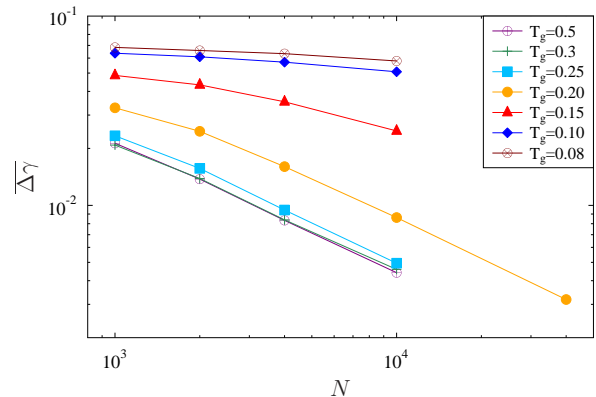}
  \caption{System size dependence of the average $\overline{\Delta\gamma}$ interval before the first plastic event occurs, with added results for $N=40000$. A $T_g$-dependent crossover in the data such as the one predicted by our toy model appears to be present.\label{fig:deltagvsN}}
 \end{center}
\end{figure}

In order to further test the plausibility of this scenario, we measured $\overline{\Delta\gamma}$ for a system size of $N=40000$ and $T_g=0.2$, which we report in Fig.~\ref{fig:deltagvsN} together with the results already shown in Fig.~\ref{fig:deltagvsNpre}. We do not simulate other values of $T_g$ as the higher ones are already in the asymptotic scaling regime, whilst the smaller ones are so far away from it that impossibly large systems are needed to see any crossover.
In any case, we are able to show that a bend in the data at low sizes is now visible, with the curve for $T_g=0.2$ apparently entering the asymptotic regime and acquiring the same slope as those for $T_g=0.3$ and $T_g=0.5$, which supports the validity of our scenario. We conclude that our data (and also those of refs.~\cite{OBBRT18,WPGLW18}) cannot unambiguously support an hypothesis of protocol dependence of the pseudogap exponent $\theta$.

\section{Conclusions and perspectives}
In summary, we have examined the scaling relation between the average interval to the first plastic event $\overline{\Delta\gamma}$ in glasses prepared from exceptionally supercooled melts with the SMC algorithm, vis-\`a-vis the proposition~\cite{OBBRT18} of a distribution of thresholds (or equivalently, local yield strains) characterized by a protocol-dependent pseudogap exponent $\theta$. Since $P(x)$ is a pdf of a locally defined observable, we employ a global measurement of the location of the first plastic event in the whole glass sample, and following~\cite{KLP10b}, we use extreme value statistics to bridge these two scales (as done also in ref.~\cite{OBBRT18}) and relate the two observables through eq.~\eqref{eq:alphavtheta}.

We find that, while a superficial analysis seems to indicate a $\theta$ exponent dependent on glass stability, this is not actually the case and in fact our data can be well explained by a simple toy model wherein the critical exponent $\theta$ is on the contrary universal by construction. The model describes well both the asymptotic and pre-asymptotic regimes of the data and accounts for the breakdown of data collapse we observe at intermediate preparation temperatures; it is also physically congruous with numerical data~\cite{18LB,WNGBSF18} which reveal that, while the DOS of quasi-localized modes (which are supposed to be the ones excited by shear strain perturbations) does indeed display a protocol dependence, this is confined to a single prefactor.
Our argument does not constitute hard proof that the exponent $\theta$ is indeed universal. One cannot exclude the presence of two effects, i.e. both a protocol dependent $\theta$ and a protocol dependent prefactor inducing pre-asymptotic corrections to scaling. Nevertheless the hypothesis appears reasonable in light of the numerics of refs.~\cite{18LB,WNGBSF18}, and testing its validity is an obvious direction for future work.

Such a test could however require system sizes too large to simulate. Intuitively, the crossover system size $N^*$ above which the asymptotic regime sets in should follow $N^* \simeq \frac{1}{c}$, $c$ being the pseudogap prefactor. If one focuses on the model glass former employed in ref.~\cite{WNGBSF18} (i.e. same interaction potential as the one we used, but in $d=3$) and identifies $c$ with the $A_4$ prefactor of the plastic DOS, the data reported in~\cite{WNGBSF18} as to its dependence on the preparation temperature seem to imply
$$
N^* \simeq \frac{1}{T_g-T_0},
$$
$T_0$ being a temperature wherein the prefactor seems to vanish; we mention that in~\cite{I18}, conversely, an exponential decrease of $A_4$ is proposed (with it never vanishing, not even at $T_g=0$) and shown to also be compatible with the data of~\cite{WNGBSF18}, but in any case, the $N^*$ crossover size grows quickly, and unfeasibly large sizes would be required in the deeply supercooled regime (the state of the art for SMC currently being around $N=10^5$ in $d=3$~\cite{WNGBSF18}), which is the regime we are interested in. In this regard, one better course of action seems to be to actually perform a local measurement of $P(x)$, as done in~\cite{PVF16,BLHGFVP18}, on glasses prepared through SMC, which should mitigate the strong finite-size effects which we have highlighted in this work.

We remark that the existence of a crossover size $N^*$ implies also the existence of an associated lengthscale,
\beq
\xi^* \simeq 1/c^{1/d},
\label{eq:lengthscale}
\eeq
$d$ being the dimensionality of space. In order to visualize it intuitively, suppose to have an amorphous solid sample whose distribution of local yield stresses corresponds to our toy model, and then to measure this distribution by dividing the solid in cells as done in~\cite{PVF16,BLHGFVP18}. The fraction of cells whose local threshold falls within the soft tail $c(T_g)x^\theta$ of the $P(x)$ is then by definition equal to the prefactor $c$, and the lengthscale $\xi^*$ is nothing but the typical distance between such cells: a system whose size is smaller than $N^* \simeq (\xi^*)^d$ will therefore not contain any of these soft cells and the extreme value argument~\eqref{eq:alphavtheta} will automatically not hold.\\
Interestingly, the definition~\eqref{eq:lengthscale} closely resembles the definition, proposed in~\cite{KSLP12}, for the static lengthscale which is posited by theories of the glass transition such as RFOT (on RFOT see for example refs.~\cite{RFOTbook,11BB,CavagnaLiq} and references therein) as being the one controlling the phenomenon of glassy slowdown~\cite{CavagnaLiq}. While the lengthscale proposed in~\cite{KSLP12} is defined in terms of the prefactor of the plastic DOS, whilst the one we study in the present work is defined in terms of the prefactor of the local distribution of thresholds (which are a priori different quantities), the possibility that the crossover lengthscale we defined here indeed corresponds to the static RFOT lengthscale is appealing, and opens the possibility of providing a bridge between the characterization of yielding and plasticity in sheared athermal glasses~\cite{PVF16,BLHGFVP18} (a problem located on the ``solid'' side of the glass transition) and the testing of theories of the glassy slowdown (a problem located on the ``liquid'' side~\cite{BBCGV08}). Notice that the link can also go the opposite way, for example, one could measure the static lengthscale in SMC glasses (as done in ref.~\cite{GKPP15}) and use it as a proxy for the prefactor of the distribution of local yield stresses of those glasses.

We finally remark that the depletion of soft excitations on lowering $c$ is a fundamentally different effect (see appendix~\ref{app:toymodel}) from the opening of a gap, such as the one found in the DOS of glasses prepared in ref.~\cite{BLW18}, wherein particles are allowed to change size in a way inspired by SMC. The study of the relation between these two effects, in particular whether or not they take place in two well-separated regimes of glass stability (and consequently, two well-separated ranges of depth in the energy landscape), is another subject for future work.

\section{Acknowledgements}
This work has been supported in part by the US-Israel BSF, the Israel Science Foundation (Program with Singapore) and the Laboratorio Congiunto ADINMAT WIS-Sapienza. We thank Eran Bouchbinder and Andrea Ninarello for useful discussions.

\appendix
\section{Estimating the Mode Coupling Transition temperature\label{app:MCT}}
We detail in this section the procedure we used to estimate the MCT transition temperature of the simulated model. The procedure is standard and follows the one used in ref.~\cite{BCNO16}.

We proceed as follows: we simulate the system (we choose $N=4000$) with simple MD dynamics (not SMC) in a range of temperatures ($T \in[0.25,0.6]$) wherein the glassy slowdown occurs, and measure its Dynamical Structure Factor~\cite{simpleliquids} $S(k,t)$, for a wavenumber $k$ corresponding to the first peak of the Static Structure Factor $S(k)$ of the melt.
\begin{figure}[htb!]
 \begin{center}
  \includegraphics[width=0.45\textwidth]{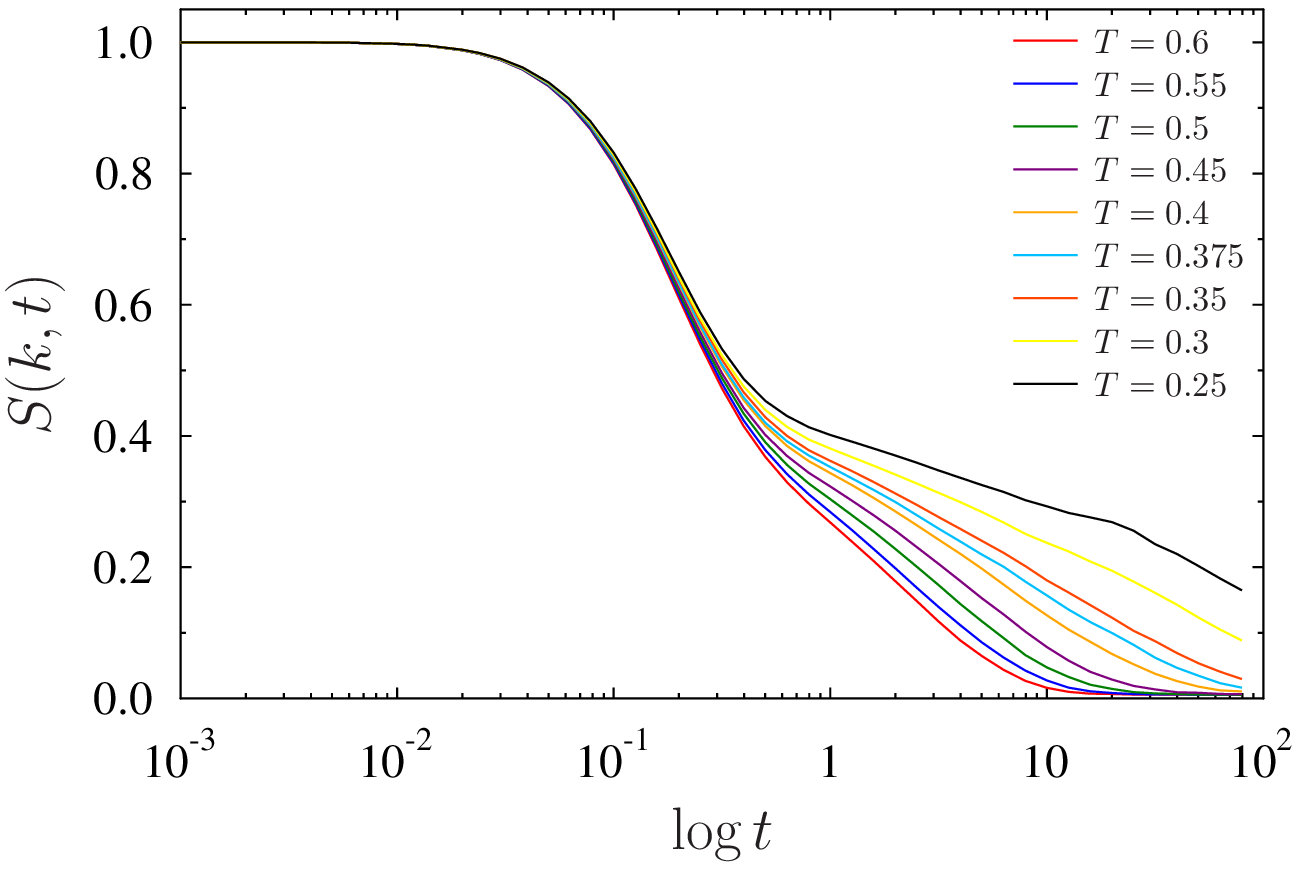}
  \includegraphics[width=0.45\textwidth]{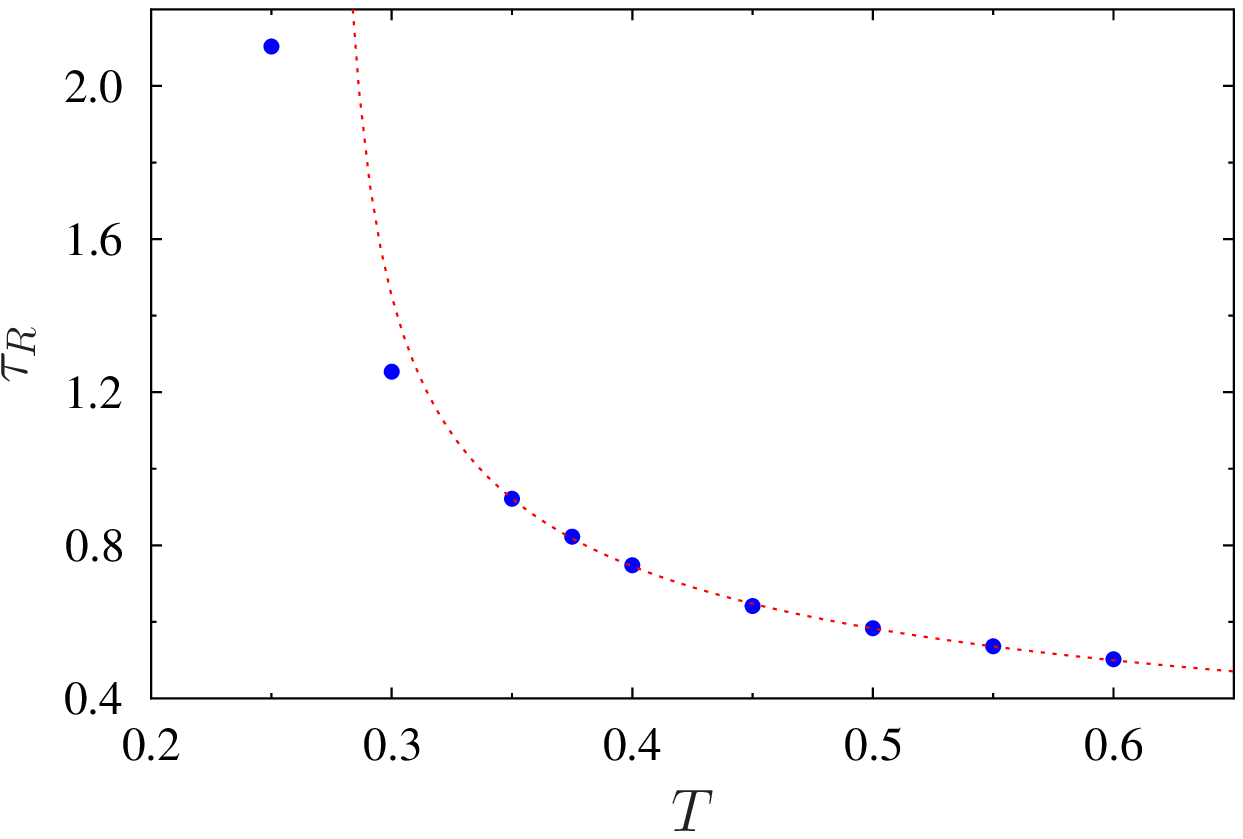}
  \caption{\label{fig:MCT} (\emph{First panel}): The Dynamical Structure Factor $S(k,t)$ of the model simulated in the main text, for various temperatures $T$ across the onset of glassy slowdown. (\emph{Second panel}) The relaxation times extracted from the decay of the $S(k,t)$, plotted against the respective temperature. The dashed red line is the best power-law fit, eq.~\eqref{eq:MCTfit} with the parameters reported in the text.}
 \end{center}
\end{figure}
In order to get rid of statistical fluctuations and obtain clean curves, we average the $S(k,t)$ over 100 initial conditions for each $T$. After the curves have been obtained (see Fig. \ref{fig:MCT}, first panel), we implicitly define a relaxation  time $\tau_R$ as
$$
S(k,\tau_R) = \frac{1}{e},
$$
which we then plot vs. $T$ in Fig.~\ref{fig:MCT}, second panel.

According to MCT~\cite{Gotze08}, the relaxation time should diverge in a power-law fashion on approaching the transition temperature $T_{MCT}$ from above. However, this does not actually happen in real glass-formers as the MCT transition is avoided and the relaxation time, although growing very fast, stays finite~\cite{CavagnaLiq}.
Nevertheless, the initial onset of the glassy slowdown is indeed well described by MCT and its power-law scaling form
\beq
\tau_R(T) = \frac{C}{(T-T_{MCT})^\gamma}
\label{eq:MCTfit}
\eeq
can be expected to apply there. We therefore choose to fit the relaxation times $\tau_R$ to the form \eqref{eq:MCTfit} in the range $T \in [0.35,0.6]$, as we deem the lowermost two $Ts$ in the second panel of Fig.~\ref{fig:MCT} to be outside the MCT regime. Furthermore, in order to avoid fitting all three parameters ($T_{MCT}$, $\gamma$, and $C$), we also fix the $\gamma$ exponent and fit only the remaining two; we then iteratively change the value of $\gamma$ and check whether the new value results in a better fit, i.e. a lower value of the reduced chisquare ($\frac{\chi^2}{n}$, $n$ being the number of degrees of freedom) parameter.\\
After a certain number of such attempts, we deem the value $\gamma=0.42$ to be the one yielding the best fit, resulting in $C=0.312\pm0.002$ and $T_{MCT} = 0.274 \pm 0.001$, which is the estimate we report in the main text. We plot the best fit in Fig.~\ref{fig:MCT}, second panel.

\section{Details of the toy model\label{app:toymodel}}
\begin{figure}[htb!]
 \begin{center}
  \includegraphics[width=0.5\textwidth]{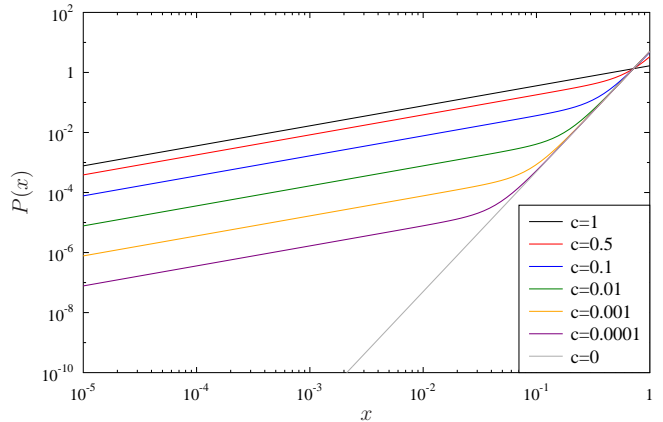}
  \caption{A cartoon showing the behavior of $P(x)$ as $c$ is varied. The extremal cases $c=0$ and $c=1$ are plotted in grey and black respectively, with the intermediate cases plotted in color.\label{fig:P(x)}}
 \end{center}
\end{figure}
The toy model for the $P(x)$ we use in the main text is a simple weighted sum of normalized power-law distributions in the interval $[0,1]$, one ``soft'' with exponent $2/3$ and one ``hard'' with exponent $4$; since the soft one dominates for low $x$, $\theta=2/3$ and constant within this model, whose only protocol dependence is assumed to be in the constant $c$.
\beq
P(x) = c\frac{5}{3}x^{2/3} + (1-c)5x^4.
\eeq
In Fig.~\ref{fig:P(x)} we show a cartoon of it for various values of $c$.
From the cartoon one can observe that, for every $c\neq0$, there is a crossover scale $\lambda$ above which the soft behavior for $x\to0$ breaks down and gives way to the ``hard'' regime, and that lowering $c$ has the effect of reducing $\lambda$. Scaling-wise one has, with generic power law exponents $\theta$ and $\eta$, $\eta>\theta$
\beq
\lambda \simeq \left(\frac{c}{1-c}\right)^{\frac{1}{\eta-\theta}},
\eeq
so in this case $1/(\eta-\theta) = 3/10$. This effect, whereupon the ``hard'' regime progressively intrudes into the range of the ``soft'', pseudogap-dominated regime, is fundamentally different from the opening of a gap in the density of local plastic thresholds.

\bibliography{LJ,bibliografia}

\end{document}